 \documentclass[final,5p,times,twocolumn]{elsarticle}

\usepackage{amssymb,amsmath,amsthm}
\usepackage{mathrsfs}
\usepackage{hyperref}
\usepackage[normalem]{ulem}

\DeclareMathOperator{\tr}{tr}
\def\widebar{\overline}
\def\mathsf{\tr}

\begin{document}

\begin{frontmatter}

\title{Bounces, turnarounds and singularities in bimetric gravity.}

 \author[a,b]{Salvatore Capozziello}
 \author[c]{Prado Mart\'{\i}n-Moruno}
\address[a]{Dipartimento di Scienze Fisiche, Universit\'a
 di Napoli ``Federico II'', Via Cinthia, I-80126 - Napoli, Italy.}
 \address[b]{INFN Sez.~di Napoli, Compl.~Univ.~di Monte S.~Angelo,
Edificio G, Via Cinthia, I-80126 - Napoli, Italy.}
\address[c]{School of Mathematics, Statistics, and Operations Research,\\
Victoria University of Wellington, PO Box 600, Wellington 6140, New Zealand.}

\begin{abstract}
In this letter, we 
 consider cosmological solutions of bimetric theory without assuming that only one metric
is coupled to gravity. We conclude that any cosmology can be described by fixing
the matter content of the space that we are not inhabiting. 
On the other hand, we show that
some conclusions can still be extracted independently of the matter content filling both spaces. 
In particular, we can conclude
the occurrence of some extremality events in one universe if we know that they take place in the other space.
\end{abstract}

\begin{keyword}
Bimetric gravity \sep Cosmology \sep Singularities

\end{keyword}

\end{frontmatter}



\section{Introduction}\label{Introduction}

The theory of bimetric gravity assumes the existence of two metric tensors interacting with each 
other \cite{Isham:1971gm}. If these metric tensors
have kinetic terms of the Einstein-Hilbert form and the equivalence principle is fulfilled, then the interaction 
between both metric fields would lead to an elegant modification of general relativity, being the existence
of the other gravitational sector only measurable by its gravitational effects.
One could wonder why this theory has not gained a renewed interest as soon as the impossibility of general 
relativity to describe our universe at astrophysical and cosmological scales (at least without introducing 
{\it ad hoc} new material components) has
been suggested by several competing approaches (see for example \cite{mauro, faraoni,tsujikawa,rew,odi,bamba,harko,olmo}).
The principal reason is that bimetric gravity generally presents a Boulware-Deser ghost \cite{Boulware:1973my}, 
which implies an instability of the theory.
Nevertheless, it has been recently shown that this undesired ghost can be discarded or controlled by 
considering  particular interactions between the metrics  \cite{Hassan:2011zd,Hassan:2011ea} (see
also \cite{nojiri} for a bigravity version of $f(R)$).

Thus, ghost-free bimetric cosmologies have been considered \cite{Volkov:2011an,vonStrauss:2011mq,Baccetti:2012ge,pilo,comelli} 
and matched to the observational data with promising results \cite{vonStrauss:2011mq,Akrami:2012vf}.
However, most of these studies are restricted to considering only a particular class of models which assume
that no material content is present in one of the spaces. In this letter we explicitly consider 
the behavior of the theory in
a cosmological scenario in the most general case, illustrating that, as it could be expected, 
the dynamics of our Universe would depend on the material content of the other space in this framework. 
As that hidden matter cannot be observed directly, this fact could be indicating a possible 
degeneracy of the theory. It seems that this degeneracy could only be cured
if the matter content of both sectors is specified from the very beginning using some argument based on fundamental
principles,  or if localized solutions are also taken into account.

Due to the complexity of analyzing the general theory, 
one can consider whether at least some information about the cosmology
of one sector can be extracted from the knowledge of the behavior of the other universe,
even without specifying the matter content of the spaces.
It is the main aim of the present letter to show that this is indeed possible regarding the occurrence of 
extremality events, as bounces, turnarounds and singularities.

\section{Cosmological solutions of the general theory}\label{degeneracy}

The action of the ghost-free bimetric gravity theory found in~\cite{Hassan:2011zd} has an interaction term
which is a function of $\gamma=\sqrt{g^{-1} f}$. That action can be re-expressed as~\cite{Baccetti:2012bk}
\begin{eqnarray}\label{action}
S&=&-\frac{1}{16\pi G}\int d^4x\sqrt{-g} \left\{R(g) + 2\,\Lambda\right\} +
\int d^4x\sqrt{-g}  \,L_\mathrm{m}
\nonumber\\
&-&\frac{\kappa}{16\pi G}\int d^4x\sqrt{-f} \left\{\widebar{R} (f)  +2\, \widebar\Lambda \right\} 
+\epsilon\int d^4x\sqrt{-f} \,\widebar L_\mathrm{m}
\nonumber\\
&+&\frac{m^2}{8\pi G}\int d^4x\sqrt{-g} L_\mathrm{int}(\gamma),
\end{eqnarray}
where the interaction Lagrangian is 
\begin{equation}\label{Lint}
 L_\mathrm{int}=\beta_1\,e_1(\gamma)+\beta_2\,e_2(\gamma)+\beta_3\,e_3(\gamma),
\end{equation}
with
\begin{eqnarray}\label{symm}
e_1(\gamma) &=&\tr[\gamma];\\
e_2(\gamma) &=&\frac{1}{2}\left(\tr[\gamma]^2-\tr[\gamma^2]\right);\\
e_3(\gamma) &=&\frac{1}{6}\left(\tr[\gamma]^3-3\tr[\gamma]\tr[\gamma^2]+2\tr[\gamma^3]\right),
\end{eqnarray}
being  elementary symmetric polynomials.
It can be noted that the effective Newton constant for the $f$-space, $\epsilon G/\kappa$, 
would be equal to that of the $g$-space only if $\epsilon=\kappa$ \cite{Baccetti:2012bk}. 
Apart from the effective Newton constant, the theory is completely symmetric under the interchange of $f$ and 
$g$ due to the properties of the elementary symmetric polynomials \cite{Hassan:2011zd,Baccetti:2012bk}.

If we consider a cosmological scenario, then, assuming that both metrics have the same sign of spatial curvature $k$, 
we can write
\begin{equation}\label{metric-g}
ds_g^2 = - dt^2 + a(t)^2 \left[ {dr^2\over 1- k r^2} + r^2 \left(d\theta^2 + \sin^2\theta\; d\phi^2\right) \right].
\end{equation}
and
\begin{equation}\label{metric-f}
ds_f^2 = - N(t)^2 \; dt^2 + b(t)^2 \left[ {dr^2\over 1- k r^2} + r^2 \left(d\theta^2 + \sin^2\theta\; d\phi^2\right) \right],
\end{equation}
where we are dismissing some special case solutions for particular values of the parameters\footnote{Those solutions are not compatible with
considering both metrics being diagonal in the same coordinate patch. That can be understood noting that the argument presented
in~\cite{Volkov:2011an} regarding the classification of solutions would be valid for any material content of both spaces
since it is based on the symmetry of the spaces.} $\beta_i$ \cite{Volkov:2011an}.
The modified Friedman equations of both spaces can be obtained by brute force from the action~(\ref{action}) 
and metrics~(\ref{metric-g}) and (\ref{metric-f}), \cite{Volkov:2011an, vonStrauss:2011mq}, or noticing that this scenario
can be described by the generalized Gordon ansatz~\cite{Baccetti:2012ge}. These are
\begin{equation}\label{Fried-g}
H_\mathrm{g}^2+\frac{k}{a^2}=\frac{m^2}{3}\rho+\frac{8\pi\,G}{3}\rho_\mathrm{m}+\frac{\Lambda}{3},
\end{equation}
and
\begin{equation}\label{Fried-f}
H_\mathrm{f}^2+\frac{k}{b^2}=\frac{m^2}{3\kappa}\widebar\rho+\frac{8\pi\epsilon G}{3\kappa}\widebar\rho_\mathrm{m}+
\frac{\widebar\Lambda}{3},
\end{equation}
with
\begin{equation}\label{rho-g}
 \rho=\frac{b}{a}\left(3\beta_1+3\beta_2\; \frac{b}{a}+\beta_3\; \frac{b^2}{a^2}\right),
\end{equation}
for the $g$-space, and
\begin{equation}\label{rho-f}
 \widebar\rho=\frac{a}{b}\left(3\beta_3+3\beta_2\; \frac{a}{b}+\beta_1\; \frac{a^2}{b^2}\right),
\end{equation}
for the $f$-space. We have defined the Hubble parameters as $H_\mathrm{g}=\dot a/a$ and $H_\mathrm{f}=\dot b/(N\,b)$,
where $\dot{} \equiv d/dt$.
The appearance of the factor $1/N(t)$ in the Hubble parameter $H_\mathrm{f}$ can be expected by noting that metric~(\ref{metric-f})
is not expressed in terms of the cosmic time of this space. We can define the cosmic time of the $f$-space as 
\begin{equation}\label{tau}
\tau (t)=\int N(t)\,dt.
\end{equation}
Thus, we are using the usual definition of the Hubble parameter also in the $f$-space, that is
$H_\mathrm{f}=b'/b$ with $'\equiv d/d\tau$.

In view of action~(\ref{action}), one can note that due to the 
diffeomorphism invariance
the matter stress energy tensor of both spaces is conserved. Thus, 
defining $w_\mathrm{m}(a)=p_\mathrm{m}/\rho_\mathrm{m}$ and $\widebar w_\mathrm{m}(b)=\widebar p_\mathrm{m}/\widebar\rho_\mathrm{m}$,
we have 
$\dot\rho_\mathrm{m}+3H_\mathrm{g}[1+w_\mathrm{m}(a)]\,\rho_\mathrm{m}=0$ and 
$\widebar\rho'_\mathrm{m}+3H_\mathrm{f}[1+\widebar w_\mathrm{m}(b)]\,\widebar\rho_\mathrm{m}=0$, which can be integrated
to obtain
\begin{equation}\label{rhom-g}
\rho_\mathrm{m}(a)=\rho_{\mathrm{m}0}\,\mathrm{exp}\left[-3\int_{a_0}^a\left[1+w_\mathrm{m}(a)\right]\frac{da}{a}\right],
\end{equation}
and
\begin{equation}\label{rhom-f}
\widebar\rho_\mathrm{m}(b)=\widebar\rho_{\mathrm{m}0}\,\mathrm{exp}\left[-3\int_{b_0}^b\left[1+\widebar w_\mathrm{m}(b)\right]\frac{db}{b}\right],
\end{equation}
respectively.
Taking into account the Bianchi identities, the stress energy tensor coming from the interaction term must be also
conserved. This leads to \cite{Volkov:2011an,vonStrauss:2011mq,Baccetti:2012ge}
\begin{equation}\label{consev}
\dot b(t)=N(t) \, \dot a(t).
\end{equation}
Therefore, the Hubble parameter of the $f$-space can be expressed as $H_\mathrm{f}=\dot a/b$, which implies that
the Friedmann equations of both spaces, (\ref{Fried-g}) and (\ref{Fried-f}), are coupled.
This fact allowed the authors of References~\cite{Volkov:2011an} and \cite{vonStrauss:2011mq} to solve the system (or to indicate
how to obtain the solutions) in the particular case that no material content is considered in the $f$-space.
In a similar way, we could, in principle, obtain the solution of the system (\ref{Fried-g}) and (\ref{Fried-f}),
taking into account Equations~(\ref{rho-g}), (\ref{rho-f}), (\ref{rhom-g}), (\ref{rhom-f}) and (\ref{consev}).
In the first place, multiplying Equation~(\ref{Fried-f}) by $b^2/a^2$, subtracting
the resulting expression from Equation (\ref{Fried-g}), inserting Equations~(\ref{rho-g}) and (\ref{rho-f}),
and simplifying the result, we obtain the following algebraic equation:
\begin{equation}\label{poli}
c_4b^4+c_3 a b^3-\frac{\widebar C}{m^2}\widebar\rho_\mathrm{m} a b^3  +c_2 a^2 b^2
+\frac{C}{m^2}\rho_\mathrm{m} a^3 b
+c_1 a^3 b-c_0 a^4=0,
\end{equation}
where $c_4=\beta_3/3$, $c_3=\beta_2-\widebar\Lambda/(3m^2)$, $c_2=\beta_1-\beta_3/\kappa$, $c_1=\Lambda/(3m^2)-\beta_2/\kappa$, 
$c_0=\beta_1/(3\kappa)$,
$C=8\pi G/3$, $\widebar C=8\pi G\epsilon/(3\kappa)$, and we are simplifying notation by assuming the dependence of
both material energy densities in their corresponding scale factors, those are $\rho_\mathrm{m}(a)$
and $\widebar\rho_\mathrm{m}(b)$.
In the second place, considering particular forms for $w_\mathrm{m}(a)$ and $\widebar w_\mathrm{m}(b)$ in Equations~(\ref{rhom-g})
and (\ref{rhom-f}), and inserting the results in Equation~(\ref{poli}),
the LHS of Equation~(\ref{poli}) can be considered as a polynomial on $b$. Thus,
once $w_\mathrm{m}(a)$ and $\widebar w_\mathrm{m}(b)$ are fixed, Equation~(\ref{poli}) can be solved to 
obtain $b$ as a function of $a$, at least in principle
(e.~g. for $\widebar w_\mathrm{m}=0$ we have a quartic equation which can be analytically solved \cite{vonStrauss:2011mq}).
In the third place, the obtained function $b(a)$ can be inserted in Equation~(\ref{rho-g}) and the result
in Equation~(\ref{Fried-g}), which could be integrated
considering again Equation~(\ref{rhom-g}).
Moreover, up to now, most studies have paid attention only to the physics of one space
(see References~\cite{Baccetti:2012ge,Baccetti:2012zz} for two interesting exceptions\footnote{Note that in
Reference~\cite{Baccetti:2012ge} a different definition of $H_\mathrm{f}$ is used.}), probably because if no material
content is considered in the other space it cannot describe an inhabited universe. Nevertheless, once we have the functions
$a(t)$ and $b(a)$,  it is straightforward to obtain $b(t)$. 
This scale factor can be more properly interpreted when it is expressed in terms of its cosmic time, $b(\tau)$, which
can also be done easily using Equations~(\ref{consev}) and (\ref{tau}).
In summary, this procedure would allow us to know the evolution of both universes once $w_\mathrm{m}(a)$ and 
$\widebar w_\mathrm{m}(b)$ are fixed.

On the other hand, it seems that one could describe any possible cosmology in the $g$-space,
i.~e.~any possible combination of $a(t)$ and $w_\mathrm{m}(a)$, by assuming a different matter content in the 
$f$-space, that is a different $\widebar w_\mathrm{m}(b)$. 
Thus, one could think that the consideration of some matter content of the $f$-space introduces a degeneracy on the theory,
and in particular on the cosmology. 
To illustrate this
different approach let us consider that $a(t)$ and $w_\mathrm{m}(a)$ are fixed and have any desired form. 
In this case we could obtain $b(t)$ from Equation~(\ref{Fried-g}) taking into account (\ref{rho-g}) and 
(\ref{rhom-g}).
Thus, from Equation~(\ref{poli}) we can obtain 
\begin{equation}\label{poliw}
\widebar\rho_\mathrm{m}(b)=\frac{m^2}{\widebar C} \left[ \frac{c_4 b}{a}+c_3+\frac{c_2 a}{b}+
\frac{C \rho_\mathrm{m} a^2}{m^2 b^2}+\frac{c_1 a^2}{b^2}-\frac{c_0 a^3}{b^3} \right] ,
\end{equation}
which can be considered as a function of $b$, since $a(b)$ can be, in principle, obtained from $b(t)$ and $a(t)$.
Therefore, taking into account Equation~(\ref{rhom-f}), this procedure implies that by choosing 
$\widebar w_\mathrm{m}(b)$ carefully in the $f$-space, 
we can obtain the desired cosmology in the $g$-space.

This second approach seems to indicate that the consideration of any matter content in the $f$-space introduces 
a degeneracy in the cosmological solutions of the theory. 
 That is, it seems that there is a loss of predictive power since any cosmology could be described
with arbitrary accuracy
by fitting a material content which cannot be observed directly.
As the model considering no material in the $f$-space can fit the observational 
data \cite{vonStrauss:2011mq,Akrami:2012vf}, one
could wonder why should we consider the possible existence of any hidden matter in the $f$-space. 
From a theoretical point of view the symmetry between the two gravitational sectors
\cite{Hassan:2011zd,Baccetti:2012zz} is one of the nicest characteristics of the theory, and postulating the absence of
matter in only one sector from the very beginning would break this symmetry. 
On the other hand, dismissing the possible existence of this matter could be an assumption stronger than considering
a particular kind of matter, if this assumption is not based in any fundamental principles.
Thus, as long as these principles are not elucidated, we are in a situation where it seems that no conclusion 
about the dynamics in both spaces can be extracted in general. 
Nevertheless, as we will show, this is not the case.

It must be pointed out that the degeneracy mentioned above is only apparent and it would not be present 
if one goes beyond the 
cosmological scenario. In particular, one could obtain the value of $\widebar w_\mathrm{m}(b)$ by fitting
the general model with future and more precise data, and test whether the predictions of the resulting theory
are fulfilled in other scenarios, for example for black hole solutions. 
Thus, the impossibility of observing directly the matter content of the other space does not imply a degeneracy
of the theory itself due to the coupling of both metrics. 
Moreover, the predictions of the theory
are, of course, different from those of general relativity even without considering perturbations, at least in
cases where both metrics are not proportional (see, e.g., Reference~\cite{Hassan:2012wr} 
for more information about the case where the metrics are proportional 
to each other).

\section{General behavior of cosmological solutions}\label{general}
We emphasize that condition~(\ref{consev}) is independent of the material content in both gravitational sectors, since it is
a consequence of the diffeomorphism invariance of action~(\ref{action}) and the particular symmetry that we are 
considering in metrics~(\ref{metric-g}) and (\ref{metric-f}). So, it seems that we could find some relation between the dynamics of both
spaces even without assuming any characteristics of the matter content. 
In particular, we can extract some information about the possible occurrence of extremality events 
in one space once we know that these events take place in the other space. 
In the first place, we express $b$ and its derivatives in terms of its cosmic time.
So, taking into account the definition~(\ref{tau}) in condition~(\ref{consev}), we have
\begin{equation}\label{deriv}
\dot a(t)=b'(\tau), 
\end{equation}
where each scale factor is derived in terms of its cosmic time. In the second place, deriving Equation~(\ref{deriv}) with respect to $t$ and using again
definition~(\ref{tau}), we obtain
\begin{equation}\label{deriv2}
\ddot a(t)=N(t)\,b''(\tau).
\end{equation}

It must be noted that 
if for a particular $t\rightarrow t_*$ we would have $N(t)\rightarrow\infty$, then, given Equation~(\ref{tau}),
that would imply $\tau\rightarrow\infty$. In this case, the region $t>t_*$ would not be in the interior\footnote{That this behavior
is indeed possible, at least in principle, can be understood considering the results in \cite{Blas:2005yk}, 
where another scenario and bimetric gravity model was considered. The authors show that the conformal diagram of one
space could be unable to accommodate all points for which the other space is defined.}
of the $f$-space. 
On the other hand, in order to have a metric with a well defined Lorentzian signature in 
the $f$-space, we must require $N(t)\neq0$. One could again interpret the possibility of attaining $N(t)=0$ as considering
a region which is not in the interior of both spaces, since $dt$ would be infinitely large for a finite $d\tau$, being, therefore,
the range of definition of $\tau$ outside that of $t$.
Thus, it would make no sense trying to extract any conclusion relating the dynamics of both spaces for a vanishing or infinite 
lapse function.

So, let us assume, for now on, that the particular times that we would refer to are in the range of definition of both spaces, 
i. e. $0<N(t)<\infty$, where we are taking both times pointing in the same direction $N(t)>0$.
Therefore, we can already conclude that if the cosmology of the $g$-space present
a bounce on a particular $t_*$, that is $\dot a(t_*)=0$ and $\ddot a(t_*)>0$, then the cosmology of the $f$-space would have
a bounce on $\tau_*=\tau(t_*)$ (which exists and is finite because we are assuming $0<N(t)<\infty$ in the
vicinity of $t_*$). This statement is obviously
also true in the opposite direction, therefore, {\it the $g$-cosmology has a bounce at $t_*$ if and only if the $f$-cosmology
has it at $\tau_*$}. It is straightforward to see that something equivalent could also be said about turnarounds, which are characterized
by $\dot a(t_*)=0$ and $\ddot a(t_*)<0$. Thus, {\it the $g$-cosmology has a turnaround at $t_*$ if and only if the $f$-cosmology
has it at $\tau_*$}.

On the other hand, the discussion of singularities is more subtle. In this case, in order to be able to extract any conclusion,
let us follow the spirit of Reference~\cite{Cattoen:2005dx} assuming that in the vicinity of a singularity the 
scale factor of the $g$-space has
a generalized power series (Puisieux series \cite{Cattoen:2006py}) expansion  (see also 
\cite{FernandezJambrina:2004yy,FernandezJambrina:2006hj}). Thus, if the $g$-cosmology is born in a 
big bang singularity at $t_*$, then for
$t\in (t_*,t_*+\delta)$ we could write
\begin{equation}\label{bang-g}
a(t)\simeq c(t-t_*)^n,
\end{equation}
where $c$ and $n$ are positive numbers (not necessarily integer numbers) and we are writing only the dominant contribution. 
Note that $0<n<1$ is needed to have
a big bang singularity, which is characterized by $a(t)\rightarrow0$, $\dot a(t)\rightarrow\infty$ and $\ddot a(t)\rightarrow-\infty$,
when $t\rightarrow t_*$. Due to Equations~(\ref{deriv}) and (\ref{deriv2}), we have $b'(\tau)\rightarrow\infty$ 
and $b''(\tau)\rightarrow-\infty$, when $\tau\rightarrow \tau_*$. So, assuming that $b(\tau)$ can also be expanded in a generalized
power series, we should have
\begin{equation}\label{bang-f}
b(\tau)\simeq d(\tau-\tau_*)^m+C,
\end{equation}
with $0<m<1$ and $d>0$, to be able to reproduce the desired divergences in $b'(\tau)$ and $b''(\tau)$. 
Thus, {\it a big bang in the $g$-universe implies also a singular origin for the $f$-universe}, although it can born
at a vanishing or non-vanishing (finite) size.
It can be noted that a similar argument would be valid with a {\it big crunch}.

Following a similar procedure, we can consider that the $g$-cosmology ends its evolution in a big rip singularity
\cite{Caldwell:2003vq}, implying
that for $t\in (t_*-\delta,t_*)$
\begin{equation}\label{rip-g}
a(t)\simeq c_1(t_*-t)^{-n_1},
\end{equation}
or in a big freeze \cite{BouhmadiLopez:2006fu}
\begin{equation}\label{freeze-g}
a(t)\simeq a_m-c_2(t_*-t)^{n_2},
\end{equation}
with $c_1,\,c_2,\,n_1>0$ and $0<n_2<1$. That is because these singularities are both characterized by $\dot a(t)\rightarrow\infty$
and $\ddot a(t)\rightarrow\infty$, for $t\rightarrow t_*$, being also $a(t)\rightarrow\infty$ for the big rip case, whereas
$a(t)\rightarrow a_m>0$ for the big freeze. Thus, these singularities differ on the value of the scale factor but 
not on the value of its derivatives. Therefore, taking into account Equations~(\ref{deriv}) and (\ref{deriv2}), 
we get in both cases
\begin{equation}\label{rip-f}
b(\tau)\simeq \frac{d}{m-1}(\tau_*-\tau)^{-m+1}+C,
\end{equation}
with $d,\,m>0$ to have $b'(\tau)\rightarrow\infty$ and $b''(\tau)\rightarrow\infty$, when $\tau\rightarrow \tau_*$. 
Nevertheless, we cannot say whether $m>1$, which would imply a big rip, or $0<m<1$, which corresponds to a big 
freeze. Anyway, we can conclude that {\it a big rip or big freeze doomsday for the $g$-universe implies also the
occurrence of a big rip or big freeze at a finite time in the $f$-space}.

In conclusion, a systematic classification of singularities at finite and infinite in both $g$ and $f$ spaces is needed in the sense discussed in  \cite{felix}.

\section{Discussion and conclusions}\label{discussion}

In this letter we have pointed out the two different approaches that one could follow when studying cosmological scenarios
in bimetric gravity. The first one entails the consideration of a particular matter content in both spaces
and completely fixes the dynamics of both universes. Whereas the second one points out that if the material content
of both spaces is not fixed by fundamental principles, then any desired universe (scale factor) filled with any material can be generated,
or reconstructed, by choosing carefully the hidden matter. 
Nevertheless, we have shown that some general conclusions about
the dynamics of both universes can be extracted without restricting the analysis to particular material contents. That is,
due to the diffeomorphism invariance, we can say that a bounce (turnaround) of one universe implies the occurrence of such
event in the other universe at a corresponding cosmic time. The presence of cosmic singularities in both spaces 
can also be concluded, since the divergent behavior of the derivatives of both scale factors in terms of their
respective cosmic times can be easily related to each other.

Finally, we want to stress that the results presented in this letter apply as long as $0<N(t)<\infty$ 
and both metrics can be expressed in the form considered in (\ref{metric-g}) and (\ref{metric-f}), that is, when both spaces are inside 
each other (the range of definition of $t$ is inside that of $\tau$ and vice versa) and the metrics can be 
taken to be diagonal in the same coordinate patch\footnote{These conditions have a similar flavor
to those considered in~\cite{Deffayet:2011rh} to study horizons in bimetric spacetimes.}.
It is known that the stress energy tensor associated to the interaction term of one of the spaces should
violate the null energy condition (if it is not saturated) when the corresponding stress energy tensor in the other space
fulfills it \cite{Baccetti:2012zz}. Thus, we expect that there would be particular cases where 
the total stress energy tensor of one space violates the null energy conditions whereas the total stress energy tensor of the other
space fulfills it, which would necessarily imply that the singularities of both cosmologies, if any, are of a different type.
Nevertheless, this behavior should correspond to a situation where the assumptions no longer hold, that is, that singularity would
take place in one space outside the range of definition of the other space.

\section*{Acknowledgments}
The authors wish to thank Matt Visser for useful comments.
SC is supported by INFN, {\it iniziativa specifica NA12}.
PMM acknowledges financial support from the Spanish Ministry of
Education through a FECYT grant, via
the postdoctoral mobility contract EX2010-0854.

\end{document}